\documentclass{aa}  
\bibpunct{(}{)}{;}{a}{}{,}

\usepackage{afterpage,natbib,lipsum}
\usepackage{graphicx}
\usepackage[usenames, dvipsnames]{color}
\usepackage{subfig}

\usepackage{multirow}
\usepackage{txfonts} 
\usepackage{hyperref} 
\usepackage{amsmath}    
\usepackage{amssymb}    
\usepackage{color}
\usepackage{rotating}
\usepackage{url}
\usepackage{ifthen}

\usepackage{bm}
\usepackage{bbold}
\usepackage{aas_macros}
\usepackage{verbatim}
\usepackage{algorithm}
\usepackage{algpseudocode}
\usepackage{enumitem}

\DeclareMathAlphabet{\mathpzc}{OT1}{pzc}{m}{it}

\newcommand{\Mpch}{{$h^{-1}$~Mpc}}

\newcommand\numberthis{\addtocounter{equation}{1}\tag{\theequation}}
\makeatletter
\newcommand{\vast}{\bBigg@{3}}
\newcommand{\Vast}{\bBigg@{4}}
\makeatother
\usepackage[dvipsnames]{xcolor}
\hypersetup{
  colorlinks   = true, 
  urlcolor     = RoyalBlue, 
  linkcolor    = RoyalBlue, 
  citecolor   = MidnightBlue 
}

\SetSymbolFont{symbols}{bold}{OMS}{cmsy}{b}{n}
\DeclareSymbolFont{bmisymbols}{OML}{cmm}{b}{it}

\usepackage{txfonts} 

\begin{document}

   \title{Explicit Bayesian treatment of unknown foreground contaminations in galaxy surveys}

   \author{Natalia Porqueres\inst{1,2}, Doogesh Kodi Ramanah\inst{3,4} \and Jens Jasche\inst{5} \and Guilhem Lavaux\inst{3,4}}

   \institute{Max-Planck-Institut f\"{u}r Astrophysik (MPA), Karl-Schwarzschild-Strasse 1 , D-85741 Garching, Germany
    \and
    Excellence Cluster Universe, Technische Universität München, Boltzmannstrasse 2, 85748 Garching, Germany 
    \and
    Sorbonne Universit\'e, CNRS, UMR 7095, Institut d'Astrophysique de Paris, 98 bis bd Arago, 75014 Paris, France             
        \and
    Sorbonne Universit\'es, Institut  Lagrange  de  Paris  (ILP),  98  bis bd Arago, 75014 Paris, France
        \and
        The Oskar Klein Centre, Department of Physics, Stockholm University, AlbaNova University Centre,
SE 106 91 Stockholm, Sweden
   }

   \date{Received 13/12/2018; accepted 14/03/2019}

 
  \abstract
   {The treatment of unknown foreground contaminations will be one of the major challenges for galaxy clustering analyses of coming decadal surveys. These data contaminations introduce erroneous large-scale effects in recovered power spectra and inferred dark matter density fields. In this work, we present an effective solution to this problem in the form of a robust likelihood designed to account for effects due to unknown foreground and target contaminations. Conceptually, this robust likelihood marginalizes over the unknown large-scale contamination amplitudes. We showcase the effectiveness of this novel likelihood via an application to a mock SDSS-III data set subject to dust extinction contamination. In order to illustrate the performance of our proposed likelihood, we infer the underlying dark-matter density field and reconstruct the matter power spectrum, being maximally agnostic about the foregrounds. The results are compared to those of an analysis with a standard Poissonian likelihood, as typically used in modern large-scale structure analyses. While the standard Poissonian analysis yields excessive power for large-scale modes and introduces an overall bias in the power spectrum, our likelihood provides unbiased estimates of the matter power spectrum over the entire range of Fourier modes considered in this work. Further, we demonstrate that our approach accurately accounts for and corrects the effects of unknown foreground contaminations when inferring three-dimensional density fields. Robust likelihood approaches, as presented in this work, will be crucial to control unknown systematic error and maximize the outcome of the decadal surveys.}


   \keywords{methods: data analysis -- methods: statistical -- galaxies: statistics -- cosmology: observations -- large-scale structure of Universe}
   
   \titlerunning{Explicit Bayesian treatment of unknown foreground contaminations}
   \authorrunning{Porqueres et al.}
   \maketitle
  
%

\section{Introduction}
\label{intro}
The next generation of galaxy surveys such as Large Synoptic Survey Telescope (LSST) \citep{lsst2008summary} or Euclid \citep{euclid2011report,euclid2016cosmology,euclid2016missiondesign} will not be limited by noise but by systematic effects. In particular, deep photometric observations will be subject to several foreground and target contamination effects, such as dust extinction, stars, and seeing \citep[e.g.][]{scranton2002analysis, ross2011ameliorating, ho2012clustering, huterer2013calibration, ho2015sloan}. 

In the past, such effects have been addressed by generating templates for such contaminations and accounting for their overall template coefficients within a Bayesian framework. \cite{leistedt2014exploiting}, for example, compiled a total set of $220$ foreground contaminations for the inference of the clustering signal of quasars in the Sloan Digital Sky Survey (SDSS-III) Baryon Oscillation Spectroscopic Survey (BOSS) \citep{bovy2012photometric}. Foreground contaminations are also dealt with in observations of the cosmic microwave background, where they are assumed to be an additive contribution to observed temperature fluctuations \citep[e.g.][]{tegmark1996method, tegmark1998measuring, hinshaw2007threeyear, eriksen2008joint, ho2015sloan, vansyngel2016semiblind, sudevan2017improved, elsner2017unbiased}. In the context of large-scale structure analyses, \cite{jasche2017foreground} presented a foreground sampling approach to account for multiplicative foreground effects which can affect the target and the number of observed objects across the sky.

All these methods rely on a sufficiently precise estimate of the map of expected foreground contaminants to be able to account for them in the statistical analysis. These approaches exploit the fact that the spatial and spectral dependence of the phenomena generating these foregrounds are well-known. But what if we are facing unknown foreground contaminations? Can we make progress in robustly recovering cosmological information from surveys subject to yet-unknown contaminations? In this work, we describe an attempt to address these questions and develop an optimal and robust likelihood to deal with such effects. The capability to account for `unknown unknowns' is also the primary motivation behind the blind method for the visibility mask reconstruction recently proposed by \cite{monaco2018blind}.

The paper is organised as follows. We outline the underlying principles of our novel likelihood in Section \ref{likelihood_formalism}, followed by a description of the numerical implementation in Section \ref{numerical_implementation}. We illustrate a specific problem in Section 4 and  subsequently assess the performance of our proposed likelihood via a comparison with a standard Poissonian likelihood in Section~\ref{results}. The key aspects of our findings are finally summarised in Section \ref{conclusion}.

\section{Robust likelihood}
\label{likelihood_formalism}

\begin{figure}
        \centering
                {\includegraphics[width=\hsize,clip=true]{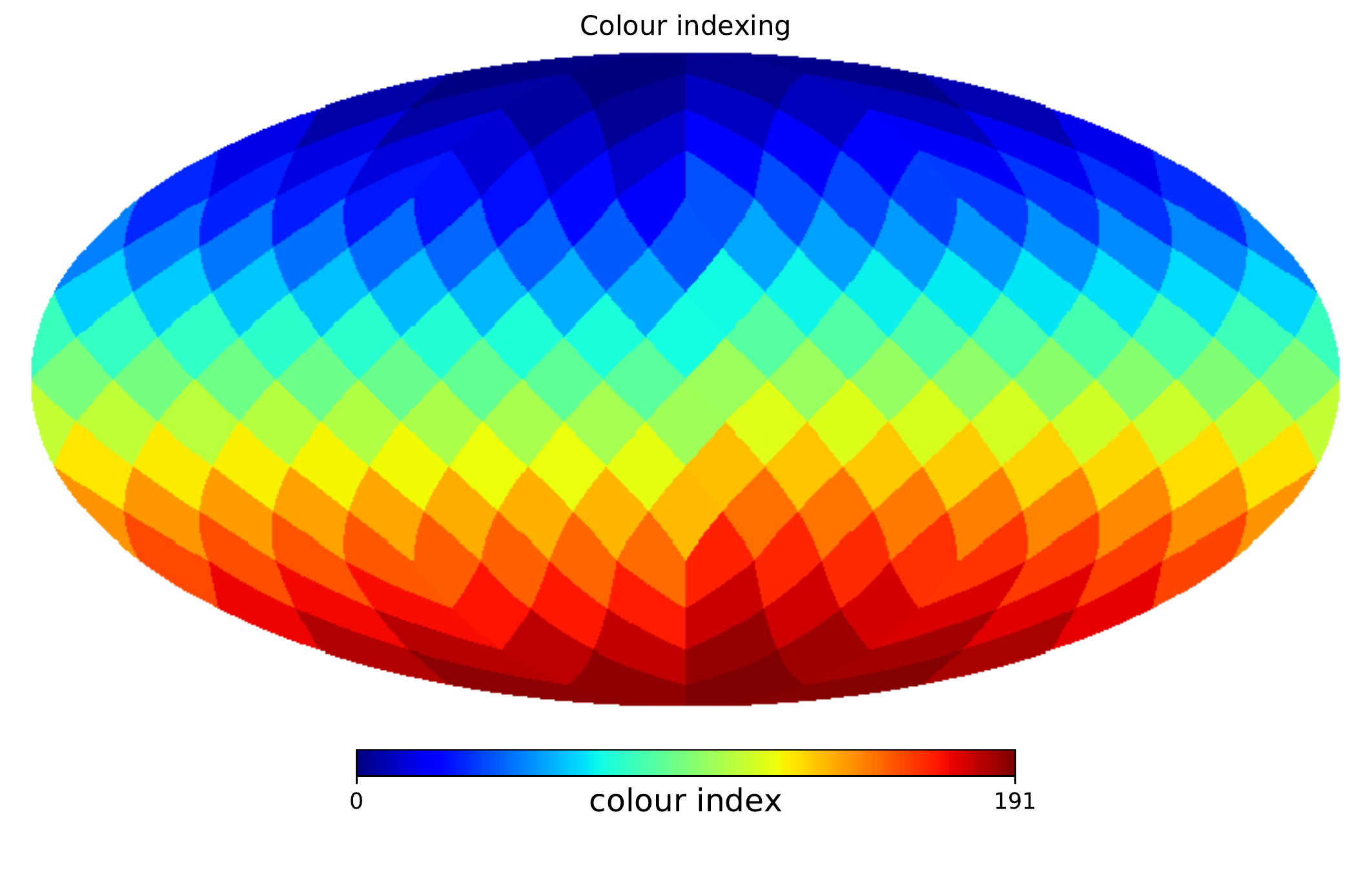}} 
        \caption{Schematic to illustrate the colour indexing of the survey elements. Colours are assigned to voxels according to patches of a given angular scale. Voxels of the same colour belong to the same patch, and this colour indexing is subsequently employed in the computation of the robust likelihood.}
        \label{fig:colouring_schematic}
\end{figure}

\begin{figure}
        \centering
                {\includegraphics[width=\hsize,clip=true]{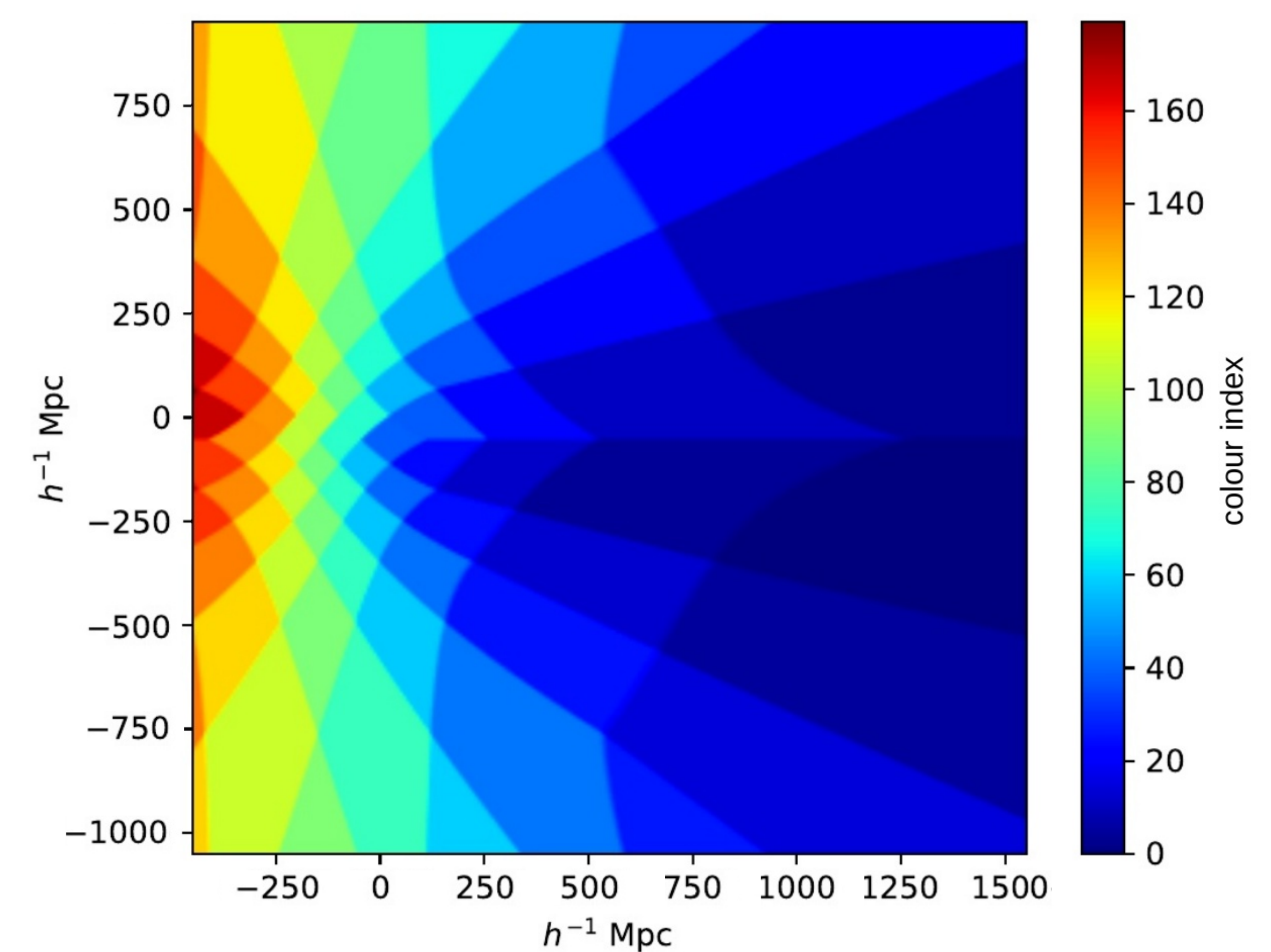}}
        \caption{Slice through the 3D coloured box. The extrusion of the colour indexing scheme (cf. Fig.~\ref{fig:colouring_schematic}) onto a 3D grid yields a collection of patches, denoted by a given colour, with a group of voxels belonging to a particular patch, to be employed in the computation of the robust likelihood. The axes indicate the comoving distances to the observer, who is located at the origin (0,0,0).}
    \label{fig:coloured_box_slice}
\end{figure}

\begin{figure}
        \centering
                {\includegraphics[width=\hsize,clip=true]{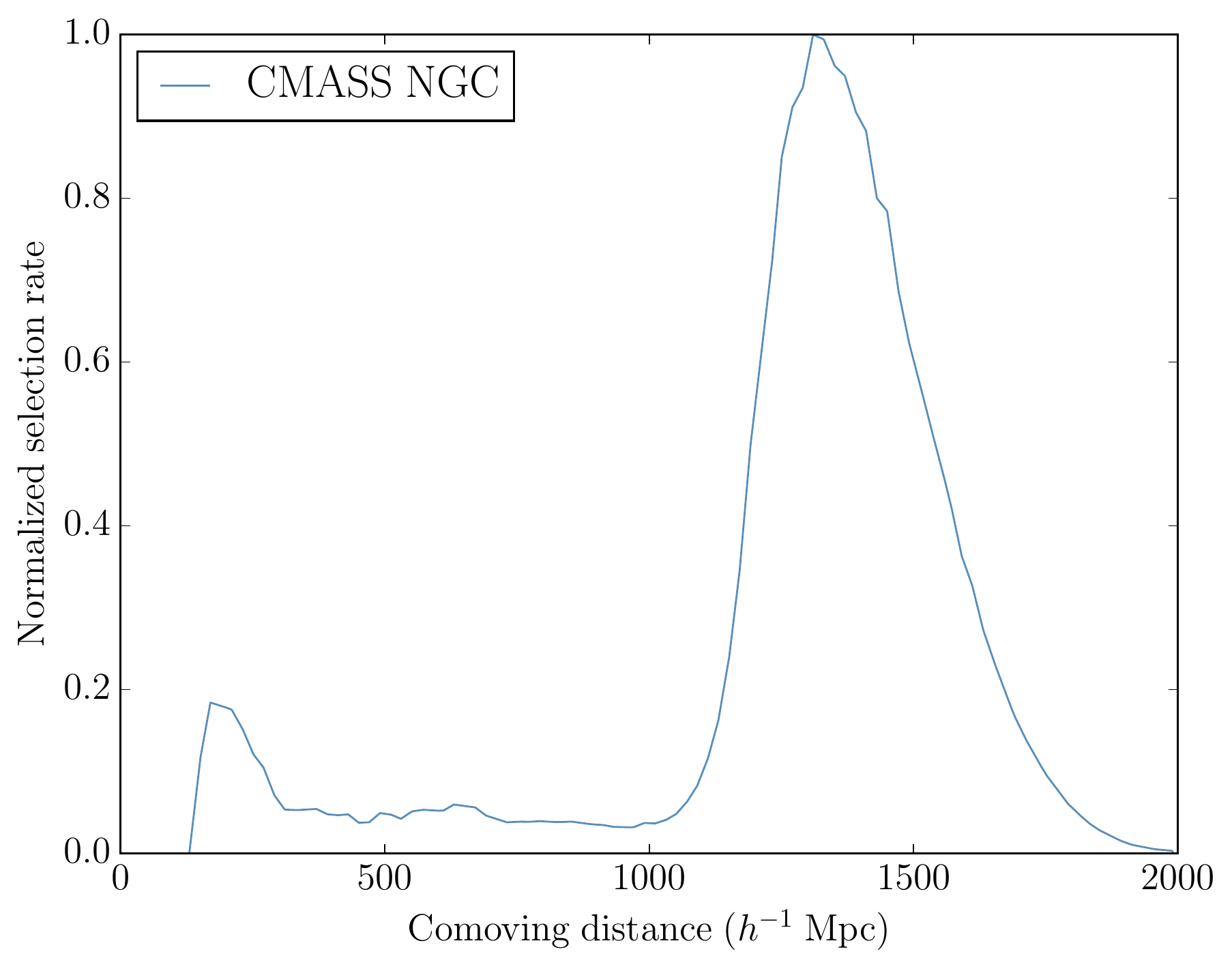}}
        \caption{Radial selection function for the CMASS (north galactic cap) survey which is used to generate the mock data to emulate features of the actual SDSS-III BOSS data.}
    \label{fig:radial_selection}
\end{figure}

\begin{figure*}
        \centering
                {\includegraphics[width=\hsize,clip=true]{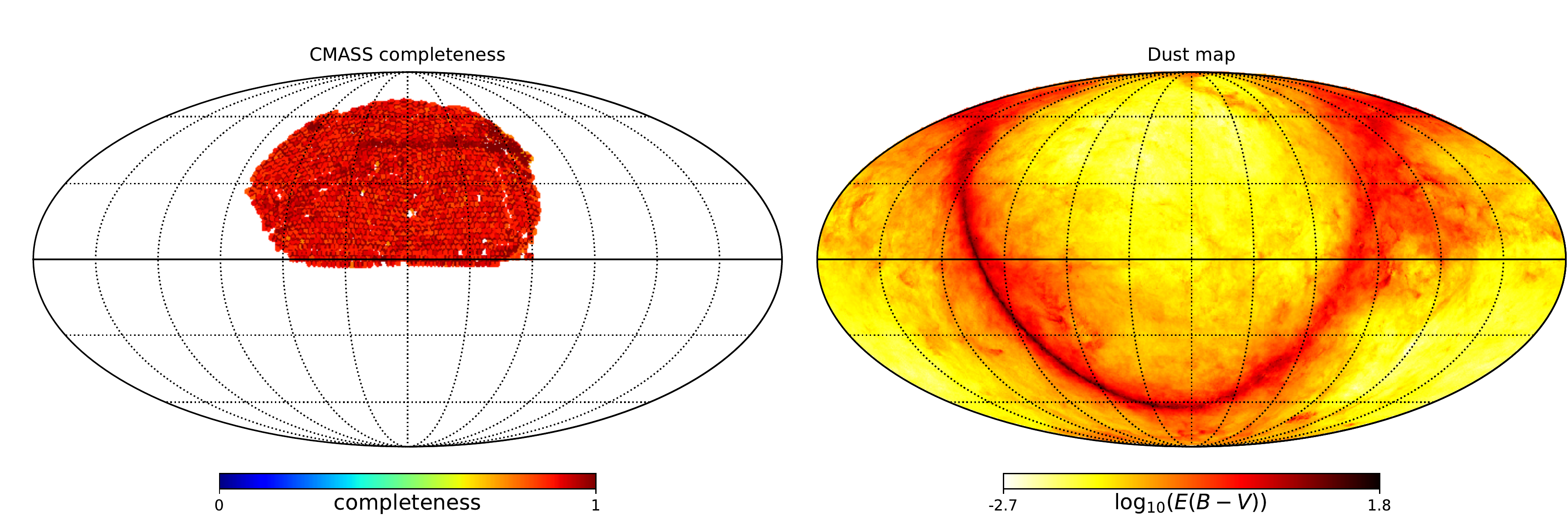}}
        \caption{Observed sky completeness ({\it left panel}) of the CMASS component of the SDSS-III survey for the north galactic cap and dust extinction map ({\it right panel}) used to generate the large-scale contamination. This reddening map has been generated from the SFD maps \citep{schlegel1998maps}.}
    \label{fig:foreground_completeness_maps}
\end{figure*}

\begin{figure*}
        \centering
                {\includegraphics[width=\hsize,clip=true]{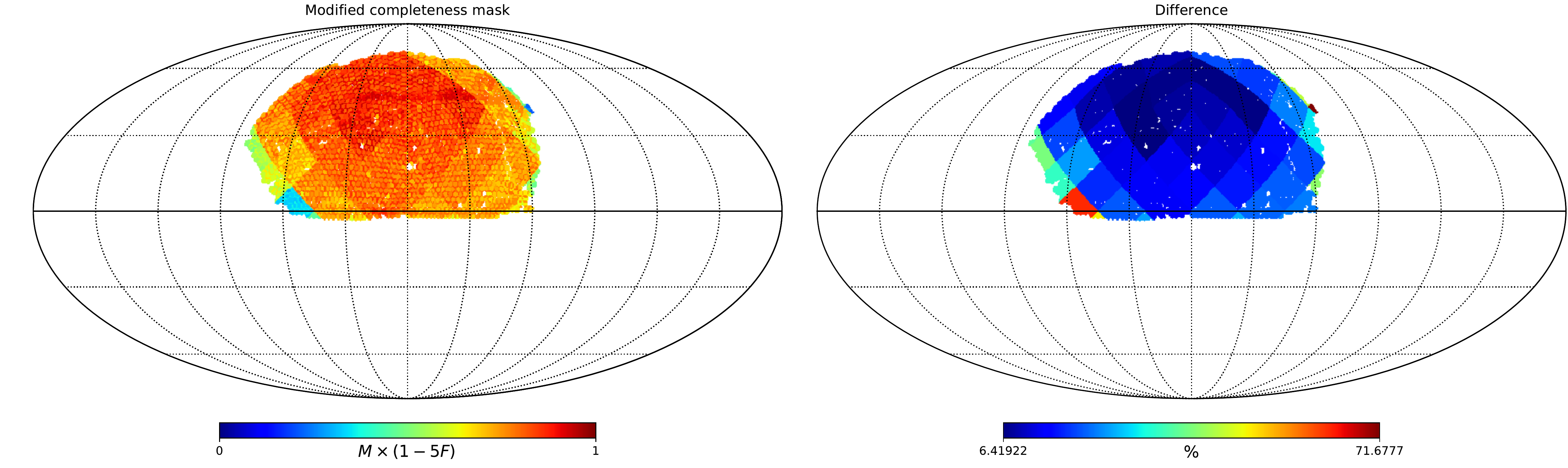}}
        \caption{Contaminated completeness mask ({\it left panel}) and percentage difference compared to the original completeness mask ({\it right panel}). The contamination is introduced by multiplying the original mask by a factor of $(1 - 5F)$ where $F$ is a foreground template, in this case, the dust extinction map downgraded to the angular resolution of the colour indexing map depicted in Fig. \ref{fig:colouring_schematic}. The factor $\alpha=5$  is chosen such that the mean contamination is 15\%, an arbitrary choice to ensure that the contaminations are significant in the completeness mask. The difference between the original and contaminated masks shows that the effect is stronger on the edges of the survey.}
    \label{fig:modified_foreground_completeness_maps}
\end{figure*}

We describe the conceptual framework for the development of the robust likelihood which constitutes the crux of this work. The standard analysis of galaxy surveys assumes that the distribution of galaxies can be described as an inhomogeneous Poisson process \citep{Layzer56,PeeblesBook,MartinezSaar03} given by 
\begin{eqnarray}
\mathcal{P}(N|\lambda) = \prod_i \frac{e^{-\lambda_i}(\lambda_i)^{N_i}}{N_i} ,
\label{eq:standard_poisson}
\end{eqnarray}
where $N_i$ is the observed number of galaxies at a given position in the sky $i$ and $\lambda_i$ is the expected number of galaxies at that position. The expected number of galaxies is related to the underlying dark-matter density field $\rho$ via 
\begin{equation}
\lambda = S\bar{N}\rho^b \exp(-\rho_g \rho^{-\epsilon}) ,
\end{equation}
where $S$ encodes the selection function and geometry of the survey, $\bar{N}$ is the mean number of galaxies in the volume, and $\{ b, \rho_g, \epsilon \}$ are the parameters of the non-linear bias model proposed by \cite{neyrinck2014halo}.

The key contribution of this work is to develop a more robust likelihood than the standard Poissonian likelihood by marginalizing over the unknown large-scale foreground contamination amplitudes. We start with the assumption that there is a large-scale foreground modulation that can be considered to have a constant amplitude over a particular group of voxels. Assuming that $A$ is the amplitude of this large-scale perturbation, we can write $\lambda_\alpha = A \bar{\lambda_\alpha}$, where the index $\alpha$ labels the voxels over which the perturbation is assumed to have constant amplitude. The likelihood consequently has the following form:

\begin{align}
\mathcal{P}(N|\bar{\lambda},A) &=       \prod_\alpha \frac{e^{-A \bar{\lambda}_\alpha}A^{N_\alpha} (\bar{\lambda}_\alpha)^{N_\alpha}}{N_\alpha} \\
&= e^{- A \sum_\alpha \bar{\lambda}_\alpha} A^{\sum_\alpha N_\alpha}  \prod_\alpha \frac{(\bar{\lambda}_\alpha)^{N_\alpha}}{N_\alpha}.
\end{align}

We can marginalize over the unknown foreground amplitude $A$ as follows:
\begin{align}
\mathcal{P}(N|\bar{\lambda}) &= \int \mathrm{d} A \; \mathcal{P} (N, A | \bar{\lambda}) \\ 
&= \int \mathrm{d} A \; \mathcal{P} (A | \bar{\lambda}) \; \mathcal{P} (N | A, \bar{\lambda}) \\ 
&= \int \mathrm{d} A \; \mathcal{P} (A) \; \mathcal{P} (N | A, \bar{\lambda}) ,
\end{align}
where, in the last step, we assumed conditional independence, $\mathcal{P} (A | \bar{\lambda}) = \mathcal{P} (A)$. This assumption is justified since the processes which generate the foregrounds are expected to be independent of the mechanisms involved in galaxy formation. As a result of this marginalization over the amplitude $A$, and using a power-law prior for $A$, $\mathcal{P} (A) = \kappa A^{-\gamma}$ where $\gamma$ is the power-law exponent and $\kappa$ is an arbitrary constant, the likelihood simplifies to:
\begin{align}
\mathcal{P}(N|\bar{\lambda}) &= \kappa \frac{\Big(\sum_{\alpha} N_\alpha\Big)!}{\Big(\sum_\beta \bar{\lambda}_\beta\Big)^{\sum_\alpha N_\alpha + 1 - \gamma}}\prod_\alpha \frac{(\bar{\lambda}_\alpha)^{N_\alpha}}{N_\alpha} \\ 
&\propto \frac{1}{\Big(\sum_\beta \bar{\lambda}_\beta\Big)^{1 - \gamma}} \prod_\alpha \Bigg(\frac{\bar{\lambda}_\alpha}{\sum_\beta \bar{\lambda}_\beta}\Bigg)^{N_\alpha}.
\label{eq:likelihood_power_law}
\end{align}

We employ a Jeffreys prior for the foreground amplitude $A$, which implies setting $\gamma = 1$. Jeffrey's prior is a solution to a measure invariant scale transformation \citep{jeffreys1946invariant} and is therefore a scale-independent prior, such that different scales have the same probability and there is no preferred scale. This scale invariant prior is optimal for inference problems involving scale measurements as this does not 
introduce any bias on a logarithmic scale. Moreover, this is especially interesting because this allows for a total cancellation of unknown amplitudes in Eq. (\ref{eq:likelihood_power_law}), resulting in the following simplified form of our augmented likelihood:
\begin{equation}
\mathcal{P}(N|\bar{\lambda}) \propto \prod_\alpha \Bigg(\frac{\bar{\lambda}_\alpha}{\sum_\beta \bar{\lambda}_\beta}\Bigg)^{N_\alpha}.
\label{eq:robust_likelihood}
\end{equation}

\section{Numerical implementation}
\label{numerical_implementation}

\begin{figure*}
        \centering
                {\includegraphics[width=\hsize,clip=true]{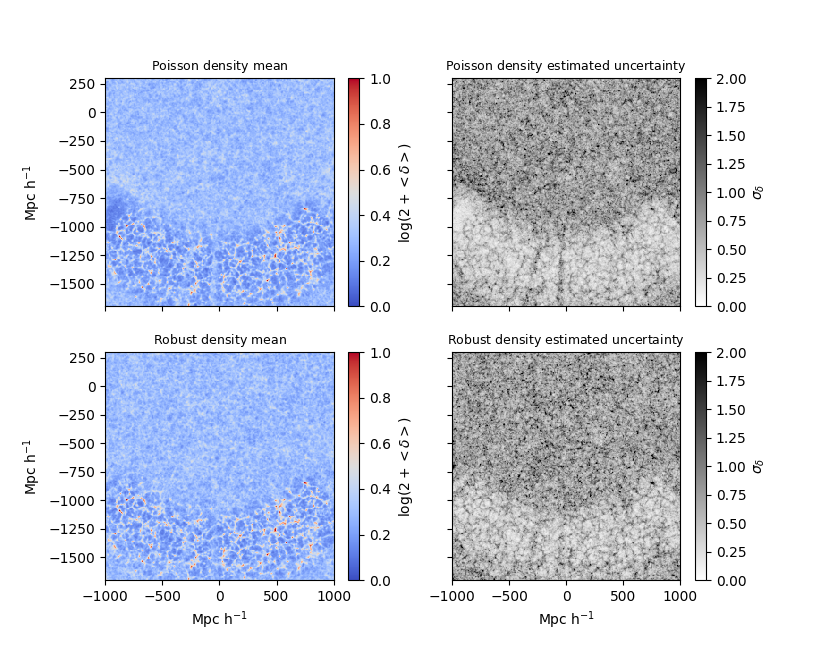}}
        \caption{Mean and standard deviation of the inferred non-linearly evolved density fields, computed from the MCMC realizations, with the same slice through the 3D fields being depicted above for both the Poissonian (upper panels) and augmented (lower panels) likelihoods. The filamentary nature of the non-linearly evolved density field can be observed in the regions constrained by the data, with the unobserved or masked regions displaying larger uncertainty, as expected. Unlike our robust data model, the standard Poissonian analysis yields some artefacts in the reconstructed density field, particularly near the edges of the survey, where the foreground contamination is stronger.} 
    \label{fig:density_correlation}
\end{figure*}

\begin{figure*}
        \centering
    \subfloat[Robust likelihood]{{\includegraphics[width=0.45\hsize,clip=true]{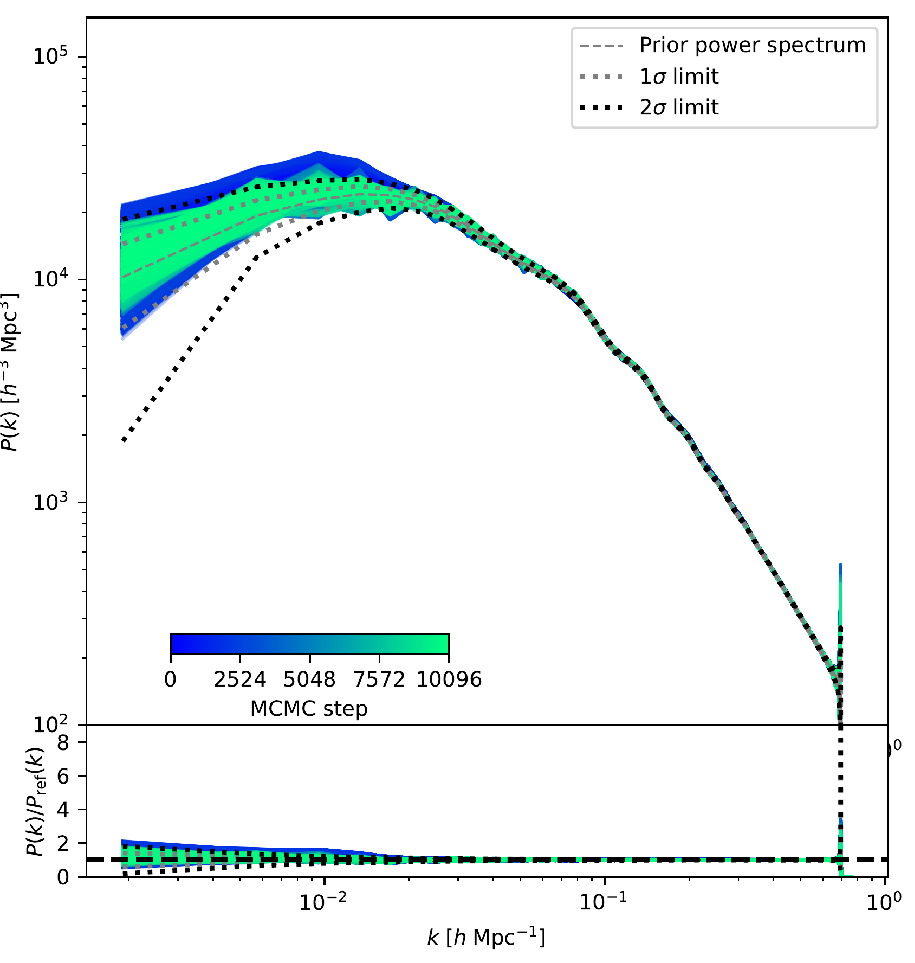} }}%
    \qquad
    \subfloat[Standard Poissonian likelihood]{{\includegraphics[width=0.45\hsize,clip=true]{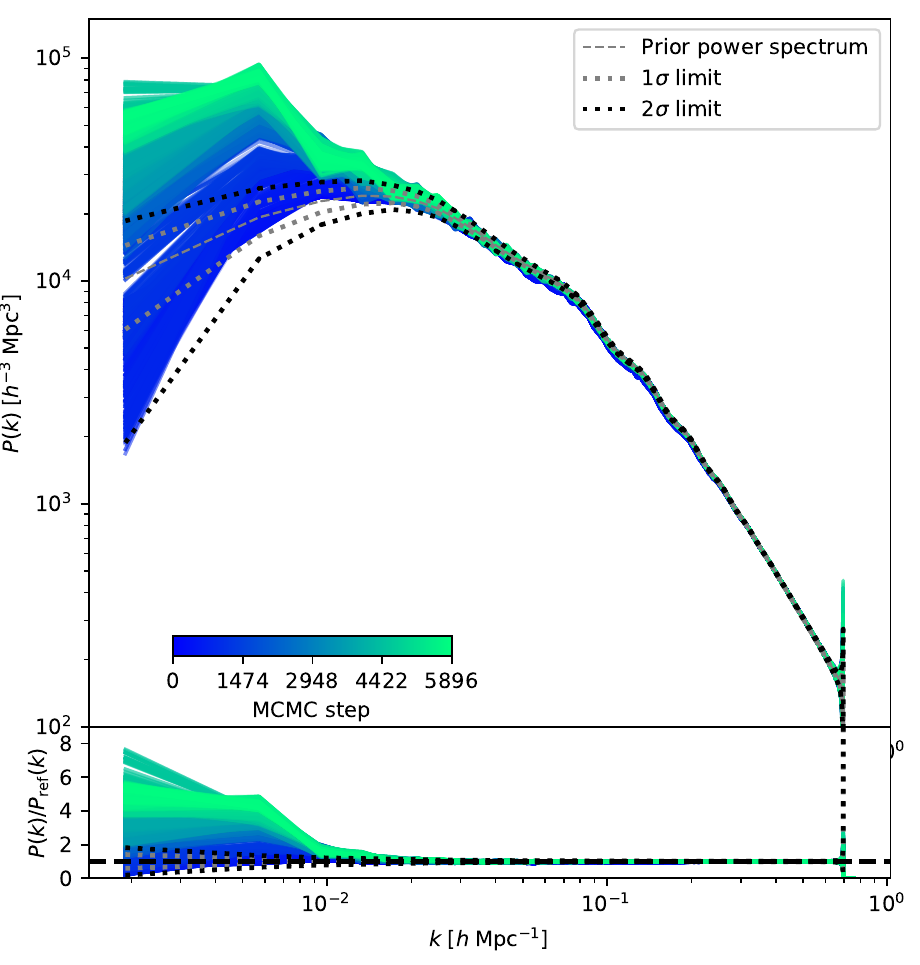} }}%
        \caption{Reconstructed power spectra from the inferred initial conditions from a \textsc{borg} analysis with unknown foreground contamination for the robust likelihood (left panel) and the Poissonian likelihood (right panel) over the full range of Fourier modes considered in this work. The $\sigma$ limit corresponds to the cosmic variance $\sigma=\sqrt[]{1/k}$. The colour scale shows the evolution of the power spectrum with the sample number. The power spectra of the individual realizations, after the initial burn-in phase, from the robust likelihood analysis possess the correct power across all scales considered, demonstrating that the foregrounds have been properly accounted for. In contrast, the standard Poissonian analysis exhibits spurious power artefacts due to the unknown foreground contaminations, yielding excessive power on these scales.}
    \label{fig:pk}
\end{figure*}

\begin{figure}
        \centering
        \includegraphics[width=\hsize,clip=true]{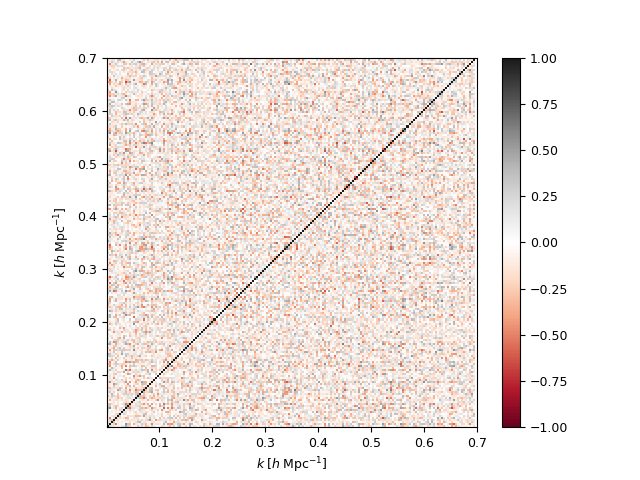}
        \caption{Correlation matrix of power spectrum amplitudes with respect to the mean value for the robust likelihood, normalized using the variance of amplitudes of the power spectrum modes. The correlation matrix shows that our augmented data model does not introduce any spurious correlation artefacts, thereby implying that it has properly accounted for the selection and foreground effects.}
    \label{fig:pk_correlation}
\end{figure}

We implement the robust likelihood in \textsc{borg} \citep[Bayesian Origin Reconstruction from Galaxies,][]{jasche2013bayesian}, a hierarchical Bayesian inference framework for the non-linear inference of large-scale structures. It encodes a physical description for non-linear dynamics via Lagrangian Perturbation Theory (LPT), resulting in a highly non-trivial Bayesian inverse problem. At the core, it employs a Hamiltonian Monte Carlo (HMC) method for the efficient sampling of a high-dimensional and non-linear parameter space of possible initial conditions at an earlier epoch, with typically $\mathcal{O}(10^7)$ free parameters, corresponding to the discretized volume elements of the observed domain. The HMC implementation is detailed in \cite{jasche2010fast} and \cite{jasche2013bayesian}. The essence of \textsc{borg} is that it incorporates the joint inference of initial conditions, and consequently the corresponding non-linearly evolved density fields and associated velocity fields, from incomplete observations. An augmented variant, \textsc{borg-pm}, employing a particle mesh model for gravitational structure formation, has recently been presented \citep{jasche2018physical}. An extension to \textsc{borg} has also been developed to constrain cosmological parameters via a novel application of the Alcock-Paczy\'nski test \citep{DKR2018altair}.

For the implementation of the robust likelihood, the HMC method that constitutes the basis of the joint sampling framework requires the negative log-likelihood and its adjoint gradient, which are given by
\begin{align*}
\Psi &\equiv -\log \mathcal{P}(N|\bar{\lambda}) \\ &= \sum_\alpha N_\alpha \log \Big( \sum_\beta \bar{\lambda}_\beta \Big) - \sum_\alpha  N_\alpha \log \bar{\lambda}_\alpha, \numberthis
\label{eq:robust_loglikelihood}
\end{align*}
and
\begin{equation}
\frac{\partial \Psi}{\partial \bar{\lambda}_\gamma}\frac{\partial \bar{\lambda}_\gamma}{\partial \rho} = \frac{\bar{\lambda}_\gamma}{\rho} \Big(b + \epsilon \rho_g \rho^{-\epsilon} \Big) \Bigg[\frac{\sum_\alpha N_\alpha}{\sum_\beta \bar{\lambda}_\beta} - \frac{N_\gamma}{\bar{\lambda}_\gamma}\Bigg].
\label{eq:robust_adjoint_gradient}
\end{equation}

The labelling of voxels with the same foreground modulation is encoded via a colour indexing scheme that groups the voxels into a collection of angular patches. This requires the construction of a sky map which is divided into regions of a given angular scale, where each region is identified by a specific colour and is stored in \texttt{HEALPix} format \citep{gorski2005healpix}, as illustrated in Fig. \ref{fig:colouring_schematic}. An extrusion of the sky map onto a 3D grid subsequently yields a 3D distribution of patches, with a particular slice of this 3D coloured grid displayed in Fig. \ref{fig:coloured_box_slice}. The collection of voxels belonging to a particular patch is employed in the computation of the robust likelihood given by Eq.~\eqref{eq:robust_loglikelihood}, where $\alpha$ corresponds to the colour index. 

This is a maximally ignorant approach to deal with unknown systematic errors where we enforce that every modulation above a given angular scale is not known. Since the colouring scheme does not depend on any foreground information, the numerical implementation of the likelihood is therefore generic. Moreover, another advantage of our approach is that the other components in our forward modelling scheme do not require any adjustments to encode this data model. However, we have not considered additive contaminations typically emanating from stars. We defer the extension of our data model to account for such additive contaminants to a future investigation.

\section{Mock generation}
\label{mock_generation}

We provide a brief description of the generation of the mock data set used to test the effectiveness of our novel likelihood, essentially based on the procedure adopted in \cite{jasche2010fast} and \cite{jasche2013bayesian}. We first generate a realization for the initial density contrast $\delta_k^{\mathrm{i}}$ from a zero-mean normal distribution with covariance corresponding to the cosmological power spectrum, such that we have a 3D Gaussian initial density field in a cubic equidistant grid with $N_{\mathrm{side}} = 256$, consisting of $256^3$ voxels, where each voxel corresponds to a discretized volume element, and comoving box length of $2000$\Mpch. This 3D distribution of initial conditions must then be scaled to a cosmological scale factor of $a_{\mathrm{init}} = 0.001$ using a cosmological growth factor $D^+ (a_{\mathrm{init}})$.

The underlying cosmological power spectrum, including baryonic acoustic oscillations, for the matter distribution is computed using the prescription described in \cite{eisenstein1998baryonic, eisenstein1999power}. We assume a standard $\Lambda$ cold dark matter ($\Lambda$CDM) cosmology with the set of cosmological parameters ($\Omega_{\mathrm{m}} = 0.3089$, $\Omega_\Lambda = 0.6911$, $\Omega_{\mathrm{b}} = 0.0486$, $h = 0.6774$, $\sigma_8 = 0.8159$, $n_{\mathrm{s}} = 0.9667$) from \cite{13planck2015}. We then employ LPT to transform the initial conditions into a non-linearly evolved field $\delta_k^{\mathrm{f}}$ at redshift $z=0$, which is subsequently constructed from the resulting particle distribution via the cloud-in-cell (CIC) method \citep[e.g.][]{hockney1988computer}.

Given the final density field $\delta_k^{\mathrm{f}}$, we generate a mock galaxy redshift catalogue subject to foreground contamination. For the test case considered in this work, we generate a data set that emulates the characteristics of the SDSS-III survey, in particular the highly structured survey geometry and selection effects. We use a numerical estimate of the radial selection function of the CMASS component of the SDSS-III survey, shown in Fig.~\ref{fig:radial_selection},  obtained by binning the corresponding distribution of tracers $N(d)$ in the CMASS sample \citep[e.g.][]{ross2017clustering}, where $d$ is the comoving distance from the observer. The CMASS radial selection function is therefore estimated from a histogram of galaxy distribution over redshift. The procedure to
construct the CMASS sky completeness is  less trivial however. We derive this CMASS mask, depicted in the left panel of Fig. \ref{fig:foreground_completeness_maps}, from the SDSS-III BOSS Data Release 12 \citep{sdss2015dr12} database by taking the ratio of spectroscopically confirmed galaxies to the target galaxies in each polygon from the mask.

In order to emulate a large-scale foreground contamination, we construct a reddening map that describes dust extinction, illustrated in the right panel of Fig.~\ref{fig:foreground_completeness_maps}. This dust template is derived from the data provided by \cite{schlegel1998maps} via straightforward interpolation, rendered in \texttt{HEALPix} format \citep{gorski2005healpix}\footnote{The construction of this template is described in more depth in Section 3 of \cite{jasche2017foreground}.}. The contamination is produced by multiplying the completeness mask of CMASS, shown in the left panel of Fig. \ref{fig:foreground_completeness_maps},  by a factor of $(1-\eta F)$, where $F$ is the foreground template rescaled to the angular resolution of the colour indexing scheme, and $\eta$ controls the amplitude of this contamination. To obtain a mean contamination of 15\% in the completeness, we arbitrarily chose $\eta = 5$  to ensure that the foreground contaminations are significant. This mean value corresponds to the average contamination per element of the sky completeness. Figure  \ref{fig:modified_foreground_completeness_maps} shows the contaminated sky completeness and the percentage difference, with the edges of the survey being more affected by the contamination due to their proximity to the galactic plane where the dust is more abundant. The mock catalogue is produced by drawing random samples from the inhomogeneous Poissonian distribution described by Eq. (\ref{eq:standard_poisson}) and using the modified completeness. 

\section{Results and discussion}
\label{results}

In this section, we discuss results obtained by applying the \textsc{borg} algorithm with the robust likelihood to contaminated mock data. We also compare the performance of our novel likelihood with that of the standard Poissonian likelihood typically employed in large-scale structure analyses. In order to test the effectiveness of our likelihood against unknown systematic errors and foreground contaminations, the algorithm is agnostic about the contamination and assumes the CMASS sky completeness depicted in the left panel of Fig. \ref{fig:foreground_completeness_maps}.

We first study the impact of the large-scale contamination on the inferred non-linearly evolved density field. To this end, we compare the ensemble mean density fields and corresponding standard deviations for the two Markov chains with the Poissonian and novel likelihoods, respectively, illustrated in the top and bottom panels of Fig. \ref{fig:density_correlation}, for a particular slice of the 3D density field. As can be deduced from the top-left panel of Fig. \ref{fig:density_correlation}, the standard Poissonian analysis results in spurious effects in the density field, particularly close to the boundaries of the survey since these are the regions that are the most affected by the dust contamination. In contrast, our novel likelihood analysis yields a homogeneous density distribution through the entire observed domain, with the filamentary nature of the present-day density field clearly seen. While we can recover well-defined structures in the observed regions, the ensemble mean density field tends towards the cosmic mean density in the masked or poorly observed regions, with the corresponding standard deviation being higher to reflect the larger uncertainty in these regions. From this visual comparison, it is evident that our novel likelihood is more robust against unknown large-scale contaminations.

From the realizations of our inferred 3D initial density field, we can reconstruct the corresponding matter power spectra and compare them to the prior cosmological power spectrum adopted for the mock generation. The top panel of Fig. \ref{fig:pk} illustrates the inferred power spectra for both likelihood analyses, with the bottom panel displaying the ratio of the {\it a posteriori} power spectra to the prior power spectrum. While the standard Poissonian analysis yields excessive power on the large scales due to the artefacts in the inferred density field, the analysis with our novel likelihood allows us to recover an unbiased power spectrum across the full range of Fourier modes.

In addition, we tested the combined effects of the foreground and unknown noise amplitudes by estimating the covariance matrix of the Fourier amplitudes of the reconstructed power spectra. As depicted in Fig. \ref{fig:pk_correlation}, our novel likelihood exhibits uncorrelated amplitudes of the Fourier modes, as expected from $\Lambda$CDM cosmology. The strong diagonal shape of the correlation matrix indicates that our proposed data model correctly accounted for any mode coupling introduced by survey geometry and foreground effects.

The above results clearly demonstrate the efficacy of our proposed likelihood in robustly dealing with unknown foreground contaminations for the inference of non-linearly evolved dark matter density fields and the underlying cosmological power spectra from deep galaxy redshift surveys. This method can be inverted to constrain foreground properties of the
contamination. The inferred dark matter density allows for galaxy catalogues to be built without contaminations. These can be compared to the observed number counts to reconstruct the foreground properties as the mismatch between the two catalogues.

\section{Summary and conclusions}
\label{conclusion}

The increasing requirement to control systematic and stochastic effects to high precision in next-generation deep galaxy surveys is one of the major challenges for the coming decade of surveys. If not accounted for, unknown foreground effects and target contaminations will yield significant erroneous artefacts and bias cosmological conclusions drawn from galaxy observations. A common spurious effect is an erroneous modulation of galaxy number counts across the sky, hindering the inference of 3D density fields and associated matter power spectra.

To address this issue, we propose a novel likelihood to implicitly and efficiently account for unknown foreground and target contaminations in surveys. We described its implementation in a framework of non-linear Bayesian inference of large-scale structures. Our proposed data model is conceptually straightforward and easy to implement. We illustrated the application of our robust likelihood to a mock data set with significant foreground contaminations and evaluated its performance via a comparison with an analysis employing a standard Poissonian likelihood to showcase the contrasting physical constraints obtained with and without the treatment of foreground contamination. We have shown that foregrounds, when unaccounted for, lead to spurious and erroneous large-scale artefacts in density fields and corresponding matter power spectra. In contrast, our novel likelihood allows us to marginalize over unknown large-angle contamination amplitudes, resulting in a homogeneous inferred density field, thereby recovering the fiducial power spectrum amplitudes. 

We are convinced that our approach will contribute to optimising the scientific returns of current and coming galaxy redshift surveys. We have demonstrated the effectiveness of our robust likelihood in the context of large-scale structure analysis. Our augmented data model remains nevertheless relevant for more general applications with other cosmological probes, with applications potentially extending even beyond the cosmological context. 

\section*{Acknowledgements}

We express our appreciation to the anonymous reviewer for his comments which helped to improve the overall quality of the manuscript. NP would like to thank Torsten En{\ss}lin for discussions and support. NP is supported by the DFG cluster of excellence ``Origin and Structure of the Universe.''\footnote{www.universe-cluster.de} This work has been done within the activities of the Domaine d’Intérêt Majeur (DIM) Astrophysique et Conditions d’Apparition de la Vie (ACAV), and received financial support from Région Ile-de-France. DKR and GL acknowledge financial support from the ILP LABEX, under reference ANR-10-LABX-63, which is financed by French state funds managed by the ANR within the Investissements d'Avenir programme under reference ANR-11-IDEX-0004-02. GL also acknowledges financial support from the ANR BIG4, under reference ANR-16-CE23-0002. This work is done within the Aquila Consortium.\footnote{\url{https://aquila-consortium.org}}

\bibliographystyle{aa.bst} 
\bibliography{robust_biblio} 
\end{document}